\documentclass[prl,aps,twocolumn,reprint,showpacs,superscriptaddress,floatfix]{revtex4-1}
\usepackage[utf8]{inputenc} 
\usepackage{graphicx}
\usepackage{amsmath}
\usepackage{amssymb}
\usepackage{amsthm} 
\usepackage{color}
 \newcommand{\be}{\begin{equation}}
\newcommand{\ee}{\end{equation}}
\newcommand{\bea}{\begin{eqnarray}}
\newcommand{\eea}{\end{eqnarray}}

\newcommand{\op}[1]{\hat{#1}} 
\newcommand{\opdag}[1]{\hat{#1}^{\dagger}} 
\newcommand{\rp}{\mathbf{r}}
\newcommand{\intV}{\int\!\! d^3 \rp} 
\newcommand{\ket}[1]{\left| #1 \right>} 

\begin{document}

\title{Supermode-Density-Wave-Polariton Condensation}
\author{Alicia J.~{Koll\'{a}r}}
\author{Alexander T.~Papageorge}
\affiliation{Department of Applied Physics, Stanford University, Stanford CA 94305}
\affiliation{E.~L.~Ginzton Laboratory, Stanford University, Stanford CA 94305}
\author{Varun D.~Vaidya}
\affiliation{E.~L.~Ginzton Laboratory, Stanford University, Stanford CA 94305}
\author{Yudan Guo}
\affiliation{E.~L.~Ginzton Laboratory, Stanford University, Stanford CA 94305}
\affiliation{Department of Physics, Stanford University, Stanford CA 94305}
\author{Jonathan Keeling}
\affiliation{SUPA, School of Physics and Astronomy, University of St Andrews, St Andrews KY16 9SS UK}
\author{Benjamin L.~Lev}
\affiliation{Department of Applied Physics, Stanford University, Stanford CA 94305}
\affiliation{E.~L.~Ginzton Laboratory, Stanford University, Stanford CA 94305}
\affiliation{Department of Physics, Stanford University, Stanford CA 94305}

\date{\today}
 \maketitle

\begin{figure*}[t!]
 \includegraphics[width=0.99\textwidth]{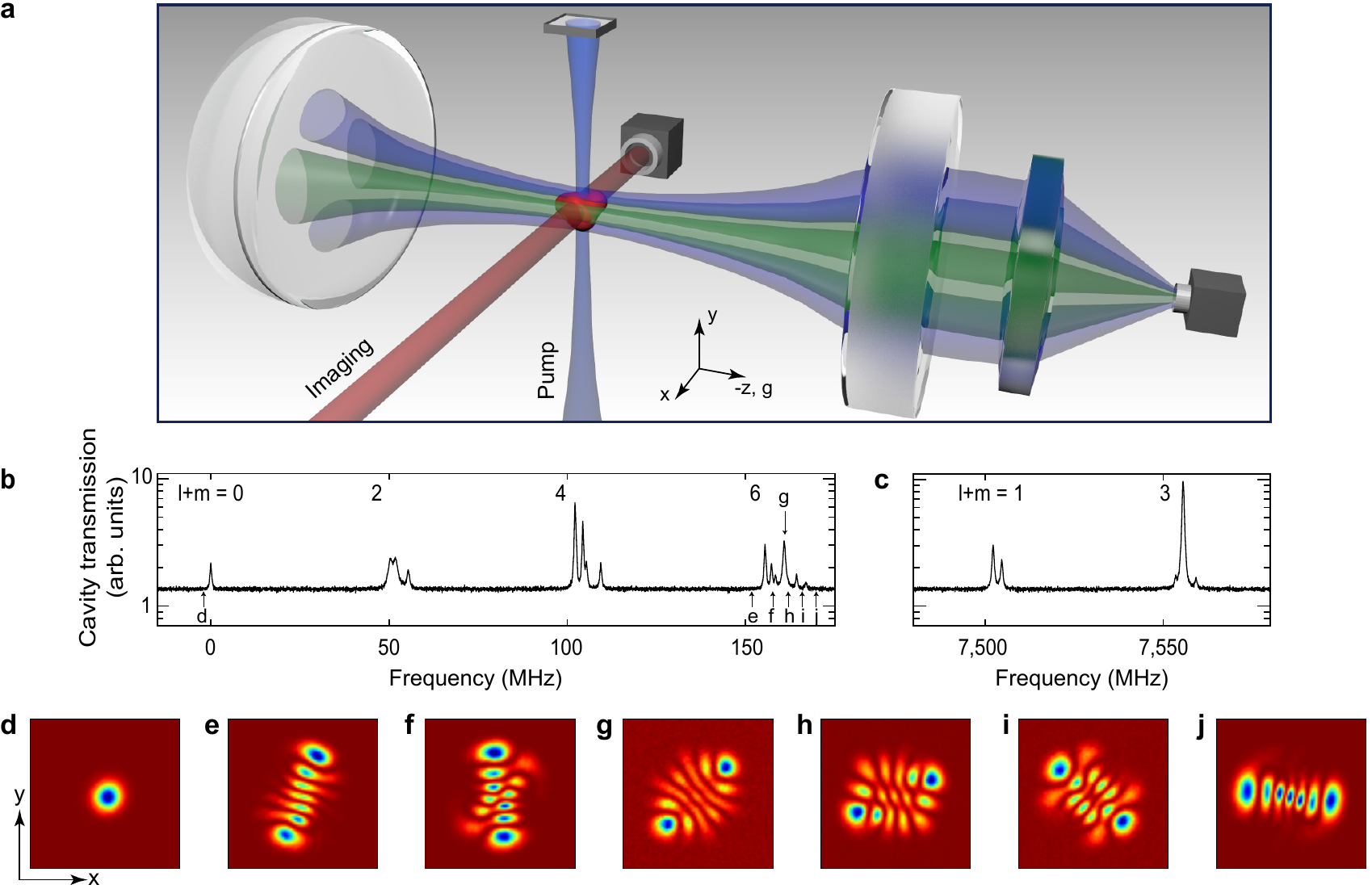}
\caption{\textbf{Experimental apparatus, cavity spectrum, and superradiant cavity emission.}  \textbf{a},  The $^{87}$Rb BEC is trapped at the center of the cavity and at the focus of the standing wave transverse pump far-detuned from  electronic transitions.  (Optical dipole trap lasers not shown.)  Detection channels include absorption imaging of atomic density and detection of cavity emission using either a single-photon counter, a photodetector, or an EMCCD camera (shown).  (b-c) Cavity transmission showing near-degenerate bare-cavity modes families (i.e., modes with $l+m=\text{const.}$) versus frequency.  These are measured with a longitudinal probe (not-shown) in the near-confocal regime for (b) even modes and (c) odd modes.   (d) Superradiant emission in the TEM$_{00}$ mode when pumped above DW-polariton condensation threshold at position \textit{d} in panel (b).  (e-j)  Superradiant photonic components of various supermode-DW-polariton condensates of the $l+m=6$ family.  Images of condensates involving higher-order modes in Extended Data Fig.~\ref{gallery}.}
\label{Apparatus}
\end{figure*}

\begin{figure}[t!]
 \includegraphics[width=0.47\textwidth]{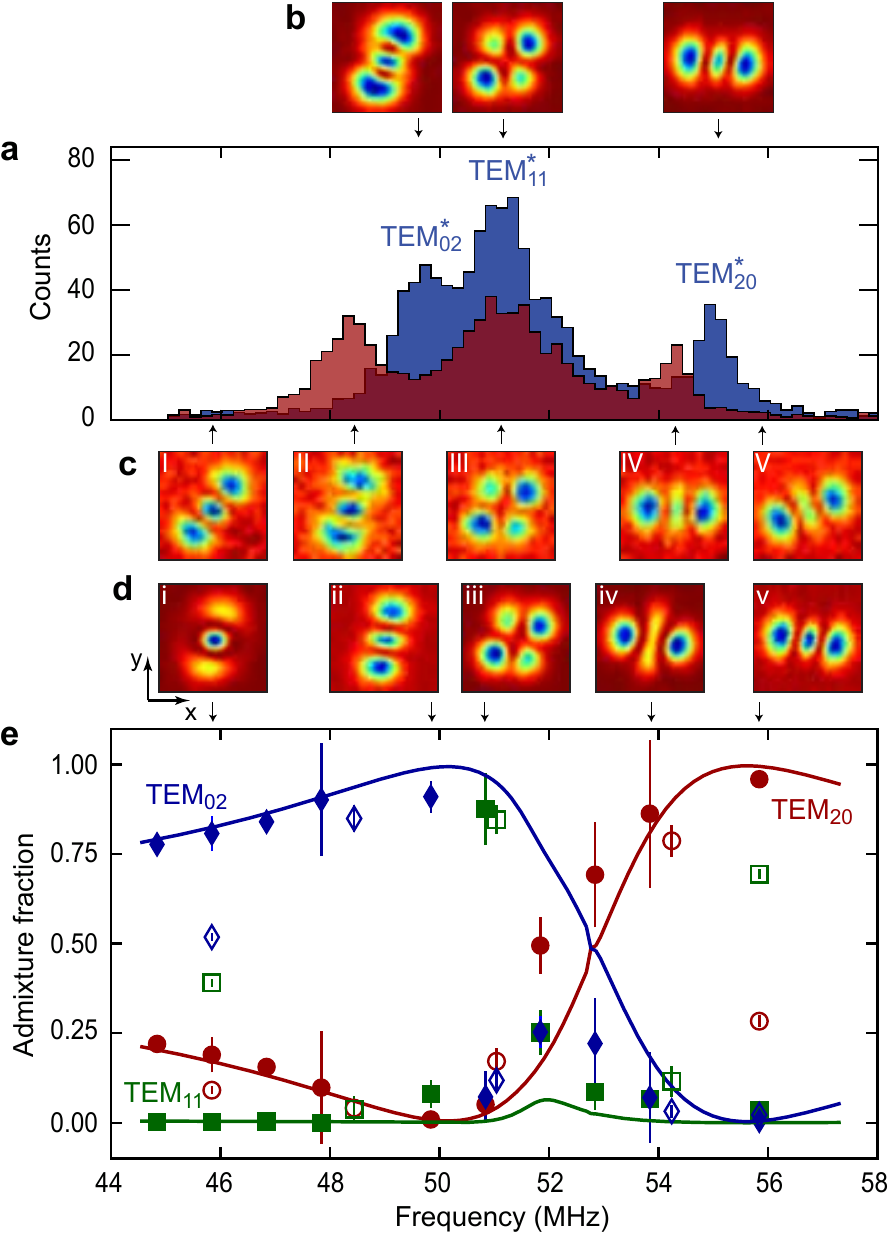}\caption{\textbf{Photonic composition.}  \textbf{a}, Mode spectrum without atoms (blue) and with atoms (red) of the $l+m=2$ family versus frequency measured on a single-photon counter with a longitudinal probe.  Frequency scale relative to TEM$_{00}$ mode in Fig.~\ref{Apparatus}b.    The three supermode peaks of the spectrum with atoms exhibit a red dispersive shift (see Methods).  Images of cavity transmission at the three bare cavity mode peaks (no atoms present) are shown in panel \textbf{b}.    The asterisks in the mode labels TEM$^*_{lm}$ indicate that these bare cavity modes are not quite ideal TEM modes due to small intrafamily mode mixing by mirror aberrations. The TEM$^*_{11}$ is unshifted due to poor overlap with centrally trapped BEC.  \textbf{c}, Images of the photonic component of the supermode-polariton dressed states, and \textbf{d}, supermode-DW-polariton condensate (pump above threshold) versus transverse-probe frequency.   The photonic composition of the near-resonance supermode-DW-polaritons \textbf{d.ii}--\textbf{d.iv} corresponds closely to that of the resonant supermodes \textbf{c.II}--\textbf{c.IV}.  However, the supermode-DW-condensates away from resonance, \textbf{d.i} and \textbf{d.v}, significantly differ from the supermodes away from resonance, \textbf{c.I} and \textbf{c.V}, demonstrating the remixing of supermodes due to DW--supermode coupling above threshold.   \textbf{e}, Admixture versus frequency of ideal cavity modes in the supermode-DW-polariton condensates (solid points) and the supermode-polaritons (open points). Error bars represent one standard error.  Solid lines are  predicted supermode-DW-polariton condensate compositions. }
\label{decomposition}
\end{figure}

\begin{figure}[t!]
\includegraphics[width=0.49\textwidth]{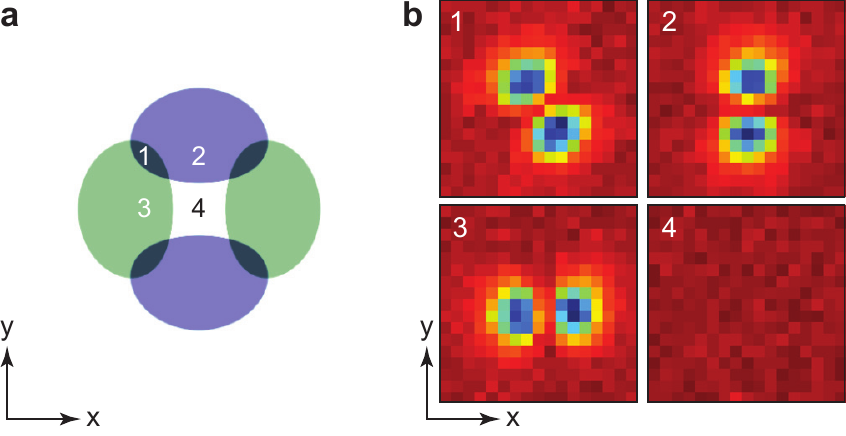}
\caption{\textbf{Dependence on BEC position.}  \textbf{a}, Schematic of the cavity transverse plane with the bare cavity TEM$_{01}$ and TEM$_{10}$ modes in blue and green, respectively.  Numbers indicate positions of the optical dipole trap confining the BEC, whose width is much smaller than cavity mode waist  (see Methods).  \textbf{b}, Moving the BEC from positions 1 through 3 changes the supermode-DW-polariton condensate composition, as may be seen by the superradiant emission in the correspondingly numbered panels.   Images taken for  same above-threshold pump power and detuning $\Delta_c = - 30$~MHz;  The astigmatic splitting of the two modes is much smaller, $2.4$~MHz.   Position 4 coincides with the $l+m=1$ family node: poor overlap lowers the coupling and threshold is not reached for this power. See Extended Data Fig.~\ref{threshold} for threshold measurements.   }
\label{position}
\end{figure}

\begin{figure*}[t!]
 \includegraphics[width=0.99\textwidth]{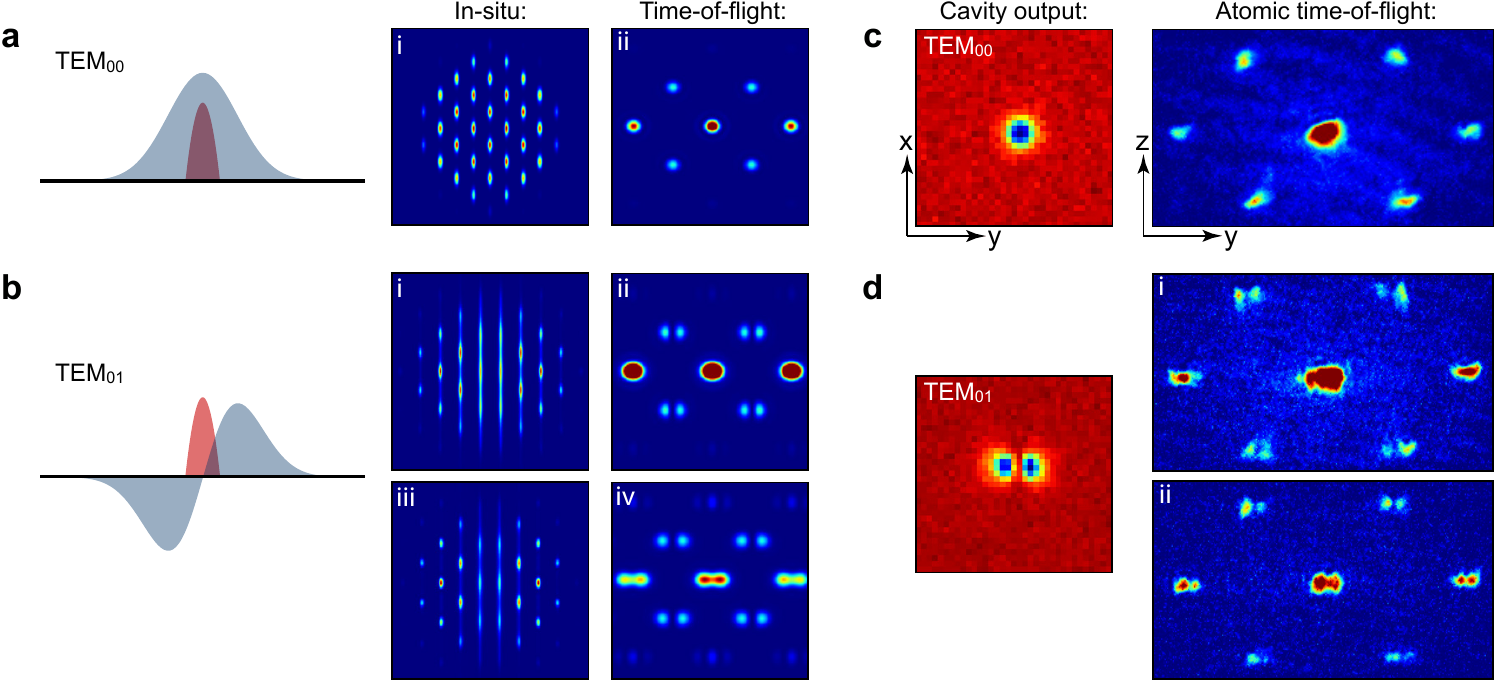}
\caption{\textbf{Structure factor measurement.} \textbf{a} and \textbf{b}, Simulation of BEC organization (see Methods).  Sketch of BEC (red) at center of \textbf{a}, TEM$_{00}$ or \textbf{b}, TEM$_{01}$ modes  (blue).   Calculated atomic density distribution shown in situ in panels \textbf{i} and \textbf{iii} and after time-of-flight expansion in panels \textbf{ii} and \textbf{iv}.   \textbf{a}, A BEC coupled to the antinode of a field organizes into a defectless checkerboard lattice above threshold, exhibiting a featureless structure factor in the experimentally observed Bragg peaks shown in the time-of-flight image in panel \textbf{c}.  (These manifestations of DW-polariton condensation are not observable in thermal gases.)   \textbf{b.i}, By contrast, BECs located at a field node  self-organize into a checkerboard lattice with a line defect at the optical node interface.   \textbf{b.ii}, This introduces a non-trivial atomic structure factor manifest as a node in the first-order Bragg peaks of the momentum distribution. \textbf{d.i}, Indeed, these nodes are clear in the time-of-flight atomic density image  for condensation into the primarily TEM$_{01}$ mode.   Image \textbf{d.ii} shows that the zero-order peak can develop structure if the system is driven with a higher pump power, as simulated in panels~\textbf{b.iii} and~\textbf{b.iv}.  This increases the coupling nonlinearity, deepening the optical potential and changing the BEC wavefunction via multiple  Bragg scattering events.   Evidence for nontrivial structure factors in supermode-DW-polartion condensates with higher-order modes is presented in Extended Data Fig.~\ref{highorderSF}.}
\label{structurefactor}
\end{figure*}

\textbf{Phase transitions, where observable properties of a many-body system change discontinuously, can occur in both open and closed systems. 
Ultracold atoms have provided an exemplary model system to demonstrate the physics of closed-system phase transitions, confirming many
theoretical models and results~\cite{Bloch:2008gl}. Our understanding of dissipative phase transitions in quantum systems is less developed, and experiments that probe this physics even less so.  By placing cold atoms in optical cavities, and inducing strong coupling between light and excitations
of the atoms, one can experimentally study phase transitions of open quantum systems.  Here we observe and study a novel form of nonequilibrium  phase transition, the condensation of supermode-density-wave-polaritons.  These polaritons are formed from a hybrid ``supermode" of cavity photons coupled to atomic density waves  of a quantum gas.  Because the cavity supports multiple photon spatial modes, and because the matter-light coupling can be comparable to the energy splitting of these modes, the composition of the supermode polariton is changed by the matter-light coupling upon condensation.  These results, found in the few-mode-degenerate cavity regime, demonstrate the potential of fully multimode cavities to exhibit physics beyond mean-field theories.  Such systems will  provide  experimental access to nontrivial phase transitions in driven dissipative quantum systems~\cite{Gopalakrishnan:2009gv,Gopalakrishnan:2010ey,Diehl2010,Chang:2013bh,Torre:2013gc,Sieberer2013} as well as enabling the studies of novel non-equilibrium spin glasses and neuromorphic computation~\cite{Gopalakrishnan:2011jx,Strack:2011hv}.}

A striking manifestation of the role quantum mechanics can play in the physics of equilibrium phase transitions is the well-known Bose-Einstein-condensation (BEC) of bosonic particles at low temperatures into a single, macroscopically populated quantum wave.   Condensation in quantum systems out of thermal equilibrium is far less understood~\cite{Diehl2010,Sieberer2013}, yet becoming experimentally relevant, especially via the study of polariton condensates.  When matter couples strongly to light, new collective modes called polaritons arise.   Condensation of these quasiparticles has been actively studied in the form  of exciton-polaritons~\cite{Kasprzak2006,Carusotto:2013gh}. However, these exist in solid-state systems, so investigating critical behavior is complicated by effects of disorder and complex phonon-mediated relaxation. Embedding the exquisite control of ultracold atoms within open quantum-optical systems provides new avenues for experimentally studying quantum fluctuation-driven transitions and quantum criticality~\cite{Gopalakrishnan:2009gv,Gopalakrishnan:2010ey,Diehl2010,Torre:2013gc,Marino:2016dp}.

We report an important step in this direction, the observation of a new form of nonequilibrium condensation, a ``supermode-density-wave-polariton" condensate. ``Supermode" refers to the photonic part of a polaritonic dressed state comprised of multiple nearly degenerate cavity modes mixed by the  intracavity atomic dielectric medium.  The supermode-polariton dressed state is dependent on  the pump-cavity detuning $\Delta_c$ as well as on the overlap of the bare-cavity modes with the BEC position and shape.    The matter component is an atomic density-wave (DW) excitation, rather than the electronic excitation of exciton-polariton condensates.   When the atoms are pumped with a laser orthogonal to the cavity axis (see Fig.~\ref{Apparatus}a), the normal modes of the atom-cavity system evolve to become supermode-DW-polaritons,  new superpositions of  supermodes mixed by the DW-fluctuations of the atoms.  Above a critical pump threshold, a supermode-DW-polariton condenses, heralded by three observables:  1)  superradiant emission of light from the cavity with the spatial pattern of one of these new supermodes; 2)  $Z_2$-symmetry breaking of the phase of the cavity field, locking to either $\varphi_0$ or $\varphi_0+\pi$ with respect to the pump phase; and 3)  organization of the BEC wavefunction into one of two checkerboard lattice configurations---each corresponding to either the $\varphi_0$ or $\varphi_0+\pi$ phase of the cavity field---but modified to  accommodate the  transverse spatial structure of the supermode.  This can result in lattice defects (matter-wave phase slips) evident as a nontrivial structure factor in an atomic time-of-flight measurement.  Observations of all these defining characteristics are presented.

With only a few modes coupled to the BEC (see Figs.~\ref{Apparatus}b and ~\ref{Apparatus}c), the supermode-DW-polariton condensate arises in a regime intermediate between  BECs coupled to a single-mode cavity and those coupled to a confocal or concentric cavity supporting many hundreds of degenerate modes.   The organized state of a matter wave coupled to a single-mode cavity has been observed, but is uniquely defined by cavity geometry alone and is best described as a DW-polariton condensate~\cite{Baumann:2010js,Nagy:2010dr,Keeling:2010by,Ritsch:2013fc}.  
Aspects of this physics have been observed in  thermal systems in which no DW-polariton condensation occurs:  e.g., phase-locking of superradiant emission~\cite{Domokos:2002bv,Black:2003bd,Baden:2014gu};
supermode emission without self-organization~\cite{Wickenbrock:2013cv}; superradiance~\cite{Greenberg:2012dx}; self-organization without superradiant emission~\cite{Labeyrie2014}.  The Bose-Hubbard model with infinite-range interactions has been studied~\cite{Klinder:2015jk,Landig:2016il}. 

The near-confocal optical cavity employed here supports families of optical modes that each lie within a small frequency bandwidth, as shown in Figs.~\ref{Apparatus}b and~\ref{Apparatus}c (see Methods).  One can observe the superradiant emission of various supermode-DW-polariton condensates by pumping at different $\Delta_c$ tuned near or within a mode family, as can be seen in Figs.~\ref{Apparatus}e through~\ref{Apparatus}j.   This is in contrast to the DW-polariton condensate of a single-mode cavity such as  shown in  Fig.~\ref{Apparatus}d:  This is not a supermode no matter the detuning $\Delta_c$ near this isolated  Gaussian mode.  The supermodes in Figs.~\ref{Apparatus}e through~\ref{Apparatus}j differ from  ideal Hermite-Gaussian modes due to three factors: 1) the bare cavity modes are themselves mixtures of ideal Hermite-Gaussian modes (due to mirror aberrations mixing modes when there is spectral overlap of modes near degeneracy)~\cite{siegman1986lasers};  2) these bare cavity modes are mixed by the dielectric atomic medium to form supermode-polaritons; 3) these dressed states are remixed by the emergent DW to form new supermodes above the polariton condensation threshold.  

Figure~\ref{decomposition} illustrates how remixed photonic components of the supermode-DW-polariton condensate can differ from the supermode-polariton dressed states.  The three bare cavity modes of the $l+m=2$ family are mixed by a BEC at the cavity center to produce the three supermode peaks.   The photonic component, shown in Fig.~\ref{decomposition}d, of the supermode-DW-polariton condensates can differ from that of the supermode-polaritons in Fig.~\ref{decomposition}c due to new supermode mixing by the macroscopically populated atomic DW above threshold.   This is most pronounced away from resonance: see, e.g., how Fig.~\ref{decomposition}c.I and~\ref{decomposition}d.i differ: the supermode-DW-polariton condensate at 46~MHz (Fig.~\ref{decomposition}d.i) is approximately 81\% TEM$_{02}$ and 19\% TEM$_{20}$, while the associated below-threshold supermode-polariton (Fig.~\ref{decomposition}c.I) is  approximately 52\% TEM$_{02}$, 9\% TEM$_{20}$ and 39\% TEM$_{11}$.  The suppression of the TEM$_{11}$ component above threshold can be understood as resulting from its poor overlap with the BEC, and thus weaker mixing with the DW mode.  A similar remixing of supermodes occurs on the blue-detuned side  at  56~MHz.  

The components closely follow the theory prediction based on a linear stability analysis for mode content at condensation threshold, except near the TEM$^*_{11}$ mode.  We believe this discrepancy is due to dynamical effects  not captured by this static stability analysis.  See Methods and Extended Data Fig.~\ref{ramprate} for discussion.  

 The position of the BEC with respect to the bare-cavity modes  affects the photonic mode composition of the supermode-DW-polariton condensate.  This is easily  observed  by moving the BEC within the transverse plane of the cavity with the pump tuned near the $l+m=1$ family (see Fig.~\ref{position}).   We pump the system with the BEC trapped by the optical dipole trap at each of the four intracavity positions illustrated in Fig.~\ref{position}a.  With the BEC trapped near either the antinode of the TEM$_{10}$ or  the TEM$_{01}$ mode, we observe superradiant emission with a spatial pattern nearly identical to these  bare-cavity modes, as shown in Fig.~\ref{position}b at positions 2 and 3.  However, a BEC at the intersection between the two modes' antinodal lobes, position 1, yields an emitted spatial pattern at 45$^\circ$ to the bare cavity mode axes.    The threshold for organization, shown in Extended Data Fig.~\ref{threshold}, is the same for a BEC at a bare-cavity antinode or between the antinodes, demonstrating that the BEC has mixed these bare modes equally and has created a basis for supermode-DW-polaritons that are rotated 45$^\circ$ from the original $l+m=1$ family eigenbasis.  
 
To complete the description of this novel non-equilibrium condensate, we report that observations of the momentum distribution in time-of-flight reveal the influence of the supermode structure on the matter wave component of the polariton condensate.    For single-mode cavities pumped near the TEM$_{00}$ mode, the atoms organise in one of two possible checkerboard patterns.
However, organization in cavities supporting higher-order modes is more complicated.  As illustrated in Figs.~\ref{structurefactor}a and~\ref{structurefactor}b,  when the BEC overlaps with a node of the cavity mode, the effective DW--cavity mode coupling changes sign across the node because the DW couples to the interference between pump field and cavity mode. This results in an organized state with a plane defect in the checkerboard lattice:  i.e., a $\pi$ phase slip, and we confirm this configuration  to be the optimal organized state via open-system simulations of the pumped BEC-cavity system.  See Fig.~\ref{structurefactor}b.i and Methods.  

Time-of-flight expansion of the atoms yields the atomic momentum distribution, and the lattice defect appears as a node in the $(k_y,k_z)=(\pm 1,\pm 1)k_R$ Bragg peaks of the expanding BEC's interference pattern; see simulation in Fig.~\ref{structurefactor}b.ii.  The node may be understood as a structure factor resulting from the low-momentum modulation of the organized atomic wavefunction caused by coupling to the transverse nodal structure of the supermode.  
The observability of a structure factor in the momentum distribution of BECs organized in orthogonally oriented and higher-order modes is shown in  Extended Data Fig.~\ref{highorderSF}.  

Stronger pumping modifies the matter wavefunction by increasing the matter-light coupling nonlinearity, as may be seen by the emerging node at the center of the zeroth-order and $(\pm 2,0)$ Bragg peaks in Fig.~\ref{structurefactor}d.ii.  This distortion of the condensate wavefunction is similar to that which happens in dilute gas BECs upon increasing interaction energy. 
Lastly, we note that these supermode-DW-polariton condensates  are observed to break the same $Z_2$-symmetry observed in single-mode DW-polariton condensates~\cite{Domokos:2002bv,Black:2003bd,Asboth:2005ho}.    Extended Data Fig.~\ref{phase} presents measurements of pump-cavity field phase locking.

The demonstration of this novel phase  in this few-mode-degenerate system paves the way for measurements of critical behavior in non-equilibrium quantum systems employing fully multimode cavity QED.  This regime can be easily realized in our existing apparatus by  tuning the cavity mirror spacing to confocality in situ~\cite{Kollar2015}.   Multimode cavity QED, in the limit of ``multimode collective ultra-strong coupling"  wherein the collective coupling is larger than the bandwidth of the degenerate modes, strongly mixing them,
should provide access to a much more exotic condensation transition.  This fluctuation-induced 1st-order Brazovskii transition is predicted to yield a superfluid smectic-like quantum liquid crystalline order of the intracavity BEC~\cite{Gopalakrishnan:2009gv,Gopalakrishnan:2010ey}.  
This opens new avenues to study  the interplay of quantum liquid crystallinity and unconventional superfluidity under controlled dimensionality and disorder~\cite{Fradkin:2012jm} as well as the study of superfluid glasses  and spin glasses~\cite{Gopalakrishnan:2009gv,Gopalakrishnan:2010ey,Habibian:2013eh}, longstanding problems in statistical mechanics~\cite{FisherHertz}.

We thank S.~Gopalakrishnan for stimulating discussions and J.~Witmer, B.~Pichler, and N.~Sarpa for early experimental assistance.   Funding was provided by the Army Research Office.   B.L.L.~acknowledges support from the David and Lucille Packard Foundation, and A.J.K.~and A.T.P.~acknowledge support the NDSEG fellowship program.  J.K.~acknowledges support from EPSRC program ``TOPNES''  (EP/I031014/1) and from the Leverhulme Trust (IAF-2014-025).

\section{Methods}
\noindent\textbf{Apparatus.} We prepare a nearly pure $^{87}$Rb Bose-Einstein condensate (BEC) of $3 \times 10^5$ atoms at the center of our cavity, confined in a crossed optical dipole trap (ODT) with trap frequencies  $\left[\omega_x,\omega_y,\omega_z\right] = 2\pi \times \left[59.1(6),88.0(9),89.4(5)\right]$~Hz. The atoms are prepared in the $\ket{F=1,m_F=-1}$ state and a $1.4$-G magnetic field is oriented along the $z$-axis. The Thomas-Fermi radii of the BEC $\left[R_x,R_y,R_z\right] = [9.8(3),8.3(2),8.3(2)]$~$\mu$m are significantly smaller than the $35$-$\mu$m waist ($1/e$ radius of the cavity field) of the TEM$_{00}$ cavity mode. The crossed ODT is formed by a pair of $1064$-nm laser beams with waists $39$ and $20$~$\mu$m intersecting at $45^{\circ}$ in the $xy$-plane. Acousto-optic modulators (AOMs) are used to stabilize the intensity of each ODT beam and control its position, allowing us to translate the BEC inside the cavity to control its overlap with the cavity modes.

The cavity is operated in a near-confocal regime in which the length $L$ is set to differ from the radius-of-curvature by $50$~$\mu$m~\cite{Kollar2015}.  The $L = 1$-cm-long cavity has a free-spectral range of $15$~GHz and a single-atom TEM$_{00}$ cooperativity of $2.5$: $g_0 = 2\pi\times 1.04$ MHz and $\kappa = 2\pi\times132$~kHz. A weak $1560$-nm laser is used to stabilize the  cavity length  using the Pound-Drever-Hall method. Additionally, light from this laser is amplified and doubled to generate $780$-nm light for the transverse pump and longitudinal probe beams. The wavelength of the locking laser is chosen to achieve a large atomic detuning of $\Delta_a = -102$~GHz between the pump and the $6$-MHz-wide D2 line of $^{87}$Rb. An electro-optic modulator (EOM) placed in the path of the locking laser beam allows us to tune the detuning $\Delta_c$ between the cavity modes and the pump or probe beams. The pump beam is polarized along $x$ and is focused down to a waist of $80$~$\mu$m at the BEC and retro-reflected to create an optical lattice oriented along $y$.

The cavity modes can be probed using a large, nearly flat longitudinal probe beam propagating along the axis of the cavity.  This probe couples to all transverse modes of the low-order families. Reference~\cite{Kollar2015} describes the design and vibration isolation of the length-adjustable cavity.

\noindent\textbf{Measurements.} The cavity output can be directed to three different detection channels.  A single-photon counting module (SPCM) can record photon numbers via a multimode fiber coupled to the multimode cavity output.   We measure a detection efficiency---from cavity output mirror to detector, including quantum efficiency and losses---of 10\% for the low-order modes discussed here. The dispersive shift data in Fig.~\ref{decomposition}a is taken for an intracavity photon number much less than one and the same $\Delta_a$ as for the rest of the data, $-102$~GHz.

Superradiant emission from the cavity is observed by monitoring the cavity output on the SPCM.  A sharp rise in intracavity photon number as the pump power is increased  heralds the condensation transition.   See Extended Data Fig.~\ref{threshold}. 

Alternatively, we can image the emission using an electron-multiplying CCD (EMCCD) camera in order to spatially resolve the transverse mode content of the cavity emission, though with no temporal resolution. All images of cavity emission are taken in a single experimental run (no averaging of shots with different BEC realizations) and with a camera integration time between $1$ and $3.3$~ms. The lower signal-to-noise in the images of Fig.~\ref{decomposition}c versus Fig.~\ref{decomposition}d is due to lower intracavity photon number.   The images in Fig.~\ref{decomposition}b with no atoms present are taken with intracavity photon number well above unity.  The pump power in Fig.~\ref{structurefactor}d.ii is 70\% larger than in Fig.~\ref{structurefactor}d.i.

The phase difference between the cavity output and pump beam can be determined by performing a heterodyne measurement with a local oscillator beam~\cite{Black:2003bd}.  The data in Extended Data Fig.~\ref{phase} has a frequency offset between pump and signal beams of $11$~MHz.  

We calibrate the Rabi frequency of the transverse pump by measuring the depth of the pump lattice through Kapitza-Dirac diffraction of the BEC. With the cavity modes detuned far-off-resonance, we pulse the pump lattice onto the BEC for a time $\Delta t$. The BEC is then released from the trap and the population $P_m(\Delta t)$ of the diffracted orders at momenta $\pm 2mk_R\hat{\mathbf{y}}$ is measured after a time-of-flight expansion. By fitting the measured $P_m(\Delta t)$ to theory~\cite{Denschlag02}, we extract a lattice depth $V_0 = \Omega^2/\Delta_a$, which allows us to determine the  Rabi frequency $\Omega$ of the pump.

The momentum distribution of the $^{87}$Rb cloud is measured by releasing the cloud from the trap and performing resonant absorption imaging after an expansion time $t_{TOF} = 17$~ms. The appearance of Bragg peaks at $|\mathbf{k}| = \sqrt{2}k_R$ in the atomic momentum distribution coincides with the onset of superradiant emission. All atomic time-of-flight images are taken in a single experimental run (no averaging of shots with different BEC realizations).  The paired atomic absorption and cavity output images  in Fig.~\ref{structurefactor}c and in Extended Data Figs.~\ref{highorderSF}a-c are each taken for the same experimental run.  The cavity output image in Fig.~\ref{structurefactor}d was taken for the same experimental run as the atomic absorption image in Fig.~\ref{structurefactor}d.i. 

We use the position dependence of the $l+m=1$ supermode-DW-polariton condensates (see Fig.~\ref{position}) to place our BEC at the center of the cavity modes. The BEC is translated in the $xy$-plane using the  AOMs of the two ODT beams. By monitoring the orientation of the superradiant emission as a function of BEC position, we are able to infer the displacement between the BEC and cavity center. Exploiting these effects allows us to position the BEC at the cavity center in both $x$ and $y$ directions to within 4~$\mu$m.

\noindent\textbf{Mode decomposition.} We analyze the transverse mode content of the polaritons in the $l+m=s$ family by decomposing the cavity field into a superposition of unit-normalized, Hermite-Gaussian modes $\Phi_{lm}(x,y;w_0,x_0,y_0)$. This is achieved by fitting the {EMCCD} image $I(x,y)$ to the function
\begin{equation}
I(x,y) = \left\lvert \sum\limits_{l+m=s}{A_{lm}e^{i\phi_{lm}} \Phi_{lm}(x,y;w_0,x_0,y_0)} \right\rvert ^2,
\end{equation} 
with the mode magnitudes $A_{lm}$ and phases $\phi_{lm}$ determined as fit parameters. The waist $w_0$ and center positions $(x_0,y_0)$ of Hermite-Gaussians are determined from an image of the TEM$_{00}$ cavity mode and are held fixed during the fit. 
Fixing the phase of the $\Phi_{s0}$ modes to $\phi_{s0}=0$  allows the fitting algorithm to converge to a local optimum. Using the fitted values of $A_{lm}$, we extract admixture fractions
\begin{equation}
f_{lm} = \frac{A_{lm}^2}{\sum_{l'm'} A_{l'm'}^2}
\end{equation}
of the $\Phi_{lm}$ mode in the cavity output.

\noindent\textbf{Model Hamiltonian.} We model the coupled dynamics of the atomic wavefunction $\op{\Psi}(\rp)$ and cavity modes $\op{a}_{\mu}$ (where $\mu = (l,m)$) with the Hamiltonian
\begin{align}
H &= \sum_j (\Delta_{\mu} + i\kappa) \opdag{a}_{\mu} \op{a}_{\mu} \nonumber\\ 
&+\intV \opdag{\Psi}(\rp) \left(\frac{\nabla^2}{2m} + V(\rp) + U \opdag{\Psi}(\rp)\op{\Psi}(\rp)\right)\op{\Psi}(\rp)\nonumber\\ 
&+ H_{ca} + H_{pa} + H_{pca}.
\label{eqn:Hfull}
\end{align}
The first term represents the evolution of the cavity modes and second is the familiar Gross-Pitaevskii Hamiltonian for a weakly-interacting BEC trapped in a harmonic potential $V(\rp)$. The atomic contact interaction is accounted for in the term proportional to $U$.  In the regime of large atomic detuning $\Delta_a$, we can neglect the excited electronic state of the atom, and the atom-light interactions may be described solely through the dispersive light shifts. The cavity-atom interaction becomes
\begin{equation}
H_{ca} = -\frac{1}{\Delta_a} \intV \opdag{\Psi}(\rp)\left(\sum_{\mu,\nu} g^*_{\mu}(\rp)g_{\nu}(\rp) \opdag{a}_{\mu}\op{a}_{\nu} \right) \op{\Psi}(\rp),
\label{eqn:Hca}
\end{equation}
where $g_{\nu}(\rp)=g_0\Phi_{\nu}(\rp)/\Phi_{00}(0)$ is the spatially dependent, single-photon (vacuum) Rabi frequency for the cavity mode $\nu$. Similarly for a pump field with a Rabi frequency $\Omega(\rp)$, the atom-pump interaction is
\begin{equation}
H_{pa} = -\frac{1}{\Delta_a} \intV \opdag{\Psi}(\rp) \Omega^*(\rp)\Omega(\rp) \op{\Psi}(\rp).
\label{eqn:Hpa}
\end{equation}
The last term of Eq.~\ref{eqn:Hfull} represents the light shift arising from the interference between the cavity and pump fields and is written as
\begin{equation}
H_{pca} =  -\frac{1}{\Delta_a} \intV \opdag{\Psi}(\rp) \left(\sum_{\mu} g^*_{\mu}(\rp)\Omega(\rp)\opdag{a}_{\mu} + \mathrm{h.c.} \right) \op{\Psi}(\rp).
\label{eqn:HPca}
\end{equation}

\noindent\textbf{Simulation of supermode composition at threshold.} 
\renewcommand{\vec}[1]{\mathbf{#1}}

To predict the location of threshold, and the nature of the supermode at that point, one may make use of a linear stability analysis, assuming a small occupation of the cavity modes and atomic density wave excitation.  For the cavity mode, this is straightforward.  For the atoms, this corresponds to assuming a condensate wavefunction
\begin{equation}
\Psi(\vec r) 
=
\psi_0(\vec r_{\perp}) \mu_0(k y)
+
\psi_1(\vec r_{\perp}) \mu_1(k y) \sqrt{2} \cos(k z),
\end{equation}
where $\mu_n(\theta)$ are $2\pi$ periodic eigenfunctions of the Mathieu equation, with eigenvalue $a_n$, that is,
$$a_n \mu_n(\theta) = [-\partial_\theta^2  + 2 q \cos(2 \theta)] \mu_n(\theta),$$
where $q=-E_{\Omega}/\omega_r$ and  $E_\Omega = \Omega^2/\Delta_a$. These Mathieu functions describe the effects of the pump beam in the $x$ direction and do not assume a weak pump lattice.  In the cavity direction, $z$, the lattice is assumed weak, and so we only consider the first two modes, i.e., $1$ and $\sqrt{2}\cos(kx)$.  Since the above expression encapsulates all effects of the longitudinal coordinate, we will suppress the label
$\perp$ on the transverse coordinates.

We must then solve coupled equations for the atomic transverse envelope functions
$\psi_n(\vec r)$ and cavity mode amplitudes $\alpha_\mu$.  To leading order in perturbation theory, the ground state envelope $\psi_0$ does not change, and so corresponds to the solution of the Gross-Pitaevskii equation:
\begin{equation}
    \mu \psi_0(\rp)
    =
    \left[
    - \frac{\nabla^2}{2m} + V(\rp)  + NU |\psi_0|^2
    \right] \psi_0(\rp),
\end{equation}
where $\mu$ is the chemical potential and $N$ the number of atoms.

Mean-field equations of motion for the $\psi_1(\rp)$ and $\alpha_{\mu} = \left\langle \op{a}_{\mu} \right\rangle$ are derived from the Hamiltonian in Eq.~\ref{eqn:Hfull},
\begin{widetext}
\begin{align}
  i \partial_t \alpha_{\mu}
  &=
   -(\Delta_{\mu} + i \kappa) \alpha_{\mu}
  - \frac{N}{2\Delta_a} 
  \int\!\! d \rp g^\ast_{\mu}(\rp) g_{\nu}(\rp) |\psi_0(\rp)|^2
   \alpha_{\nu}
  -
   \frac{N \Omega O(q)}{\Delta_a}
  \int\!\! d \rp g^\ast_{\mu}(\rp) (\psi_0(\rp) \psi_1^\ast(\rp)
    + \text{H.c.}),
  \\  
  i \partial_t \psi_1(\rp)
    &=
    \left[
    \omega_0(q) 
    - \frac{\nabla^2}{2m} + V(\rp)  + 2 NU |\psi_0|^2
    \right] \psi_1(\rp)
    + N U \psi_0(\rp)^2 \psi_1^\ast(\rp)
    -
    \frac{\Omega}{\Delta_a} 
    \left(\sum_\mu \alpha_\mu g_\mu(\rp) + \text{H.c.} \right) 
    \psi_0(\rp),
\end{align}
\end{widetext}
where we have taken $\Omega(\vec{r}) = \Omega \cos(kx)$. The spatial dependence of the pump enters through the overlap $O(q) = \left< \sqrt{2}\cos(\theta) \mu_0(\theta) \mu_1(\theta)\right>$, of the first two Mathieu functions due to  the cross pump-cavity light field potential.  The energy scale $\omega_0(q) = \omega_r(1 + a_1(q) - a_0(q))$ corresponds to the effective recoil in pump and cavity directions, allowing for the possibility of a deep pump lattice.
For a shallow lattice these functions become $1$ and $2 \omega_r$ respectively.

From these linearised equations, we may then determine when supermode-DW-polariton condensation occurs, by identifying the point at which the linearised fluctuations become unstable.  There is some subtlety to this point, discussed further below.
Calculating the growth/decay rates of linearised fluctuations is straightforward, corresponding to an eigenvalue equation.  Because there are anomalous coupling terms (i.e., because
$\alpha_\mu$ depends on both $\psi_1(\rp)$ and $\psi_1^\ast(\rp)$, and vice versa) one must use the Bogoliubov--de Gennes parametrisation, i.e., write
$\alpha_\mu(t) = \alpha_{\mu,+}e^{i\lambda t} + \alpha_{\mu,-}e^{-i \lambda t}$ and similarly for $\psi_1(\rp)$. 
It is convenient to resolve the function $\psi_1(\rp)$ onto some set of basis states.  We use the harmonic oscillator basis states, giving a particularly simple result in the limit $U \to 0$.

With the basis noted above, the eigenvalue problem is given by
$\text{Det}[A - \lambda 1]=0$ where the matrix $A$ can be written in the block form in terms of $(\alpha_+, \alpha_-, \psi_{1+}, \psi_{1-})$
blocks:
\begin{widetext}
\begin{equation}
  \label{eq:Amatrix}
  A = 
  \begin{pmatrix}
    i \kappa + \vec{\Delta}_c + \frac{E_0}{2} \vec{M} & 0 & 0 & 0 \\
    0 & i \kappa - \vec{\Delta}_c - \frac{E_0}{2} \vec{M}^\ast & 0 & 0 \\
    0 &  & - \vec{\Delta}_{dw}(q) - 2\vec{W} & -\vec{W} \\
    0 & 0  & \vec{W}  &  \vec{\Delta}_{dw}(q) + 2 \vec{W}
  \end{pmatrix}
  +
  \sqrt{E_0 E_\Omega} O(q)
  \begin{pmatrix}
    0 & 0    &  \vec{Q}^\ast &  \vec{Q}^\ast \\
    0 & 0    & - \vec{Q} & - \vec{Q} \\
    \vec{Q}^T &  \vec{Q}^\dagger     & 0 & 0 \\
    - \vec{Q}^T & - \vec{Q}^\dagger  &    0  &  0
  \end{pmatrix}.
\end{equation}
\end{widetext}
In this expression, the various block matrices are as follows: The matrix $\vec{\Delta}_c$ is a diagonal matrix consisting of the detuning between the pump laser and each cavity mode.  $\vec{\Delta}_{dw}(q)$ is similarly a diagonal matrix describing the energy difference between a given atomic transverse mode function and the ground state mode function.  This is a function of $q$ as it also includes the energy $\omega_0(q)$ associated with the different scattering states.   The matrices $\vec{W}$ denote the effect of atom-atom interactions, corresponding to the overlap between two atomic modes and the atomic ground state density.  That is, they describe scattering off the condensate causing transitions between modes. The matrices $\vec{M}$ denote the dielectric shift due to the atoms, corresponding to the overlap between two cavity modes and the atomic ground state density.  (The matrices $\vec{M}$ and $\vec{W}$ differ in general because the cavity beam waist does not match the harmonic oscillator length of the atoms.)  The matrix $\vec{Q}$ denotes atom-cavity scattering and involves the overlap of the atomic ground state mode function with a given excited mode and a given cavity mode.  The energy scale is $E_0 = N(g_0^2/\Delta_a)$ for cavity beam waist $w$.

On solving Eq.~\eqref{eq:Amatrix}, one finds that while there is a threshold for instability when  $\vec{\Delta}_c<0$ of nearby modes,  there is always an unstable eigenvalue as soon
as $\vec{\Delta}_c>0$ for any mode (i.e., the pump is blue-detuned of any mode).  However, the growth rate of this instability varies widely with parameters.  At low pump powers, the timescale for growth is very long (i.e., seconds)~\cite{Bhaseen:2012cq}.  As pump strength increases, there is a sharp threshold where two eigenvalues of Eq.~\eqref{eq:Amatrix} cross, demarcating a transition to a state that rapidly orders (timescale of microseconds).  The curves in Figure 2 correspond to finding the eigenvector (i.e., mode composition) of this mode which becomes rapidly unstable.

\noindent\textbf{Structure factor simulation.} The calculated density and momentum distributions presented in Fig.~\ref{structurefactor} are evaluated by numerically integrating the  mean-field equations of motion in 3D for the atomic wavefunction and cavity mode. Using the Hamiltonian in Eq.~\ref{eqn:Hfull}, we derive equations of motion for the atomic wavefunction and cavity field under a mean-field approximation where $\left\langle \op{a}_{\mu} \right\rangle = \alpha_{\mu}$ and $\langle \op{\Psi}(\rp) \rangle = \sqrt{N}\psi(\rp)$. This gives us the coupled differential equations,
\begin{align}
&i\partial_t \alpha_{\mu} = -(\Delta_{\mu}+i\kappa)\alpha_{\mu}  \\
&-\frac{N}{\Delta_a} \intV \psi(\rp,t) \left(g^*_{\mu}(\rp)\sum_{\nu} g_{\nu}(\rp)\alpha_{\nu}\right)\psi(\rp,t)  \nonumber\\
&-\frac{N}{\Delta_a} \intV \psi(\rp,t)  g^*_{\mu}(\rp)\Omega(\rp)  \psi(\rp,t), \nonumber
\end{align}
\begin{align}
&i\partial_t \psi(\rp,t) = \left(-\frac{\nabla^2}{2m} + V(\rp) + NU |\psi(\rp,t)|^2\right)\psi(\rp,t) \\ 
&- \frac{1}{\Delta_a} \left( \sum_{\mu,\nu} g^*_{\mu}(\rp)g_{\nu}(\rp) \alpha_{\mu}\alpha_{\nu} \right)\psi(\rp,t) \nonumber\\ 
&- \frac{1}{\Delta_a}\left( \sum_{\mu} g^*_{\mu}(\rp)\Omega(\rp)\alpha_{\mu} + \mathrm{h.c.} \right)\psi(\rp,t) .\nonumber
\end{align}
Furthermore we adiabatically eliminate the cavity field $\alpha_{\mu}$ under the assumption that it equilibrates on a timescale much faster than the atomic motion. To simulate the behavior presented in Fig.~\ref{structurefactor}, we restrict  $\alpha_{\mu}$ to a single mode, either TEM$_{00}$ or TEM$_{10}$, and numerically integrate the equations of motion. The initial atomic wavefunction is set to the Thomas-Fermi distribution associated with our ODT parameters, and the initial cavity field is set to $\alpha_{\mu}(0) = 0$. The strength of the pump field is increased linearly in time from $0$ to $\Omega(\rp,t)$ to simulate the transverse pumping of our cavity. The in situ density distributions shown in Fig.~\ref{structurefactor}a are $|\psi(\rp,t)|^2$. The momentum distributions are obtained by a Fourier transform of the in situ atomic wavefunction.

\onecolumngrid

\begin{figure}[h]
\includegraphics[width=0.75\textwidth]{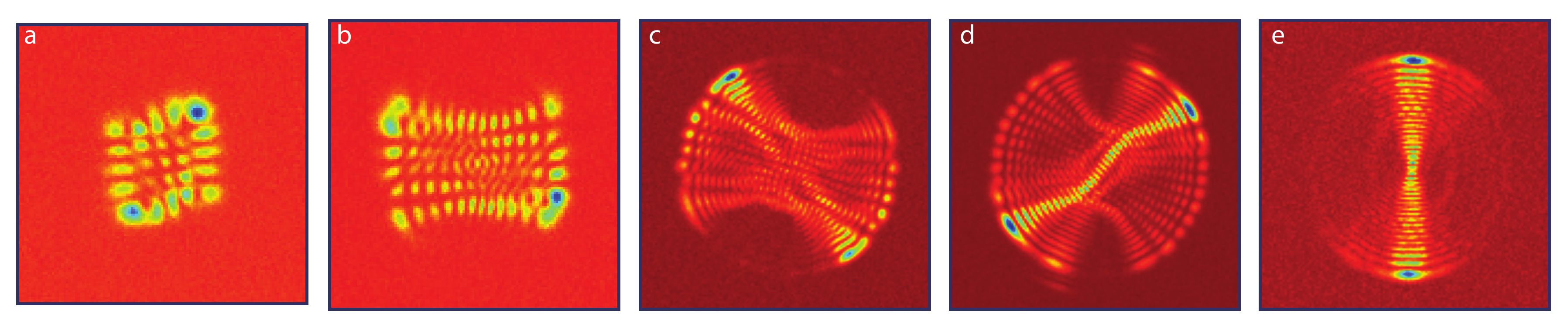}
\caption{\textbf{Gallery of supermode-DW-polariton condensates.}  \textbf{a}--\textbf{e}, Photonic components of various high-order supermode-DW-polariton condensates.  Emission clipped at large radius by limited aperture of camera imaging optics at this magnification.}
\label{gallery}
\end{figure}

\begin{figure}[b]
 \includegraphics[width=0.75\textwidth]{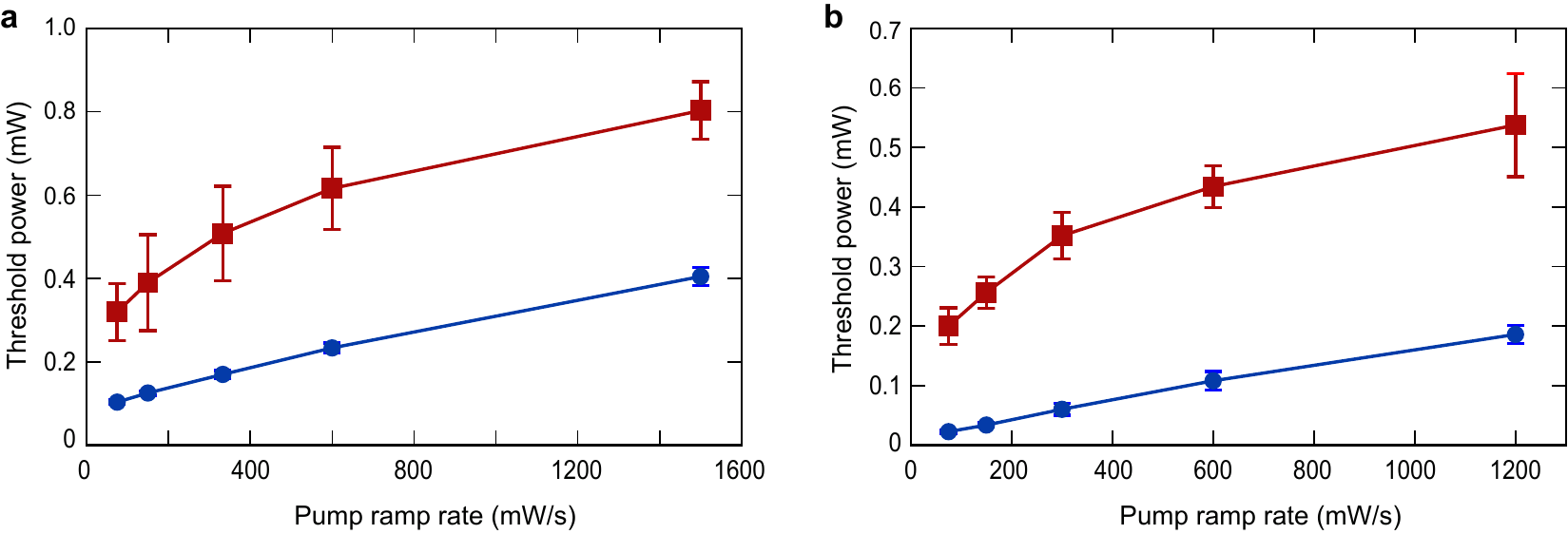}
\caption{\textbf{Threshold as a function of transverse pump power ramp rate.} \textbf{a}, A comparison between the thresholds for the $l+m = 0$ (blue) and $l+m = 1$ (red) modes for a BEC placed at the center of the cavity. In both cases the BEC is pumped at a large detuning of $\Delta_c = -30$~MHz. The $l+m = 0$ mode has an antinode at the center of the cavity, while the $l+m = 1$ mode has a node.  The overall higher threshold of the odd mode is indicative of the worse overlap between the atomic wavefunction and the node of the optical mode.   The greater rate of threshold increase with pump power ramp rate indicates that the atoms need more time to adapt to the sign-flip of the odd mode than can be provided  during the timescale of the faster ramp.  Higher threshold pump power is needed to compensate.  A consequence of this can be seen in the slow turn-on of superradiance in the data of Fig.~\ref{threshold}b.  This dynamical effect hints that motion of the atoms following the turn-on of the cavity light may in some cases lead to effects beyond the linear stability analysis presented in the Methods section, which produced the green curve in Fig.~\ref{decomposition}e.   \textbf{b}, An analogous behavior is seen in the $l+m = 2$ family, for the TEM$_{02}$ (blue) and TEM$_{11}$ (red) modes. The former has an antinode at the cavity center, while the latter, while still an even-parity mode, has a node at the cavity center. The thresholds were measured by pumping at a detuning $\Delta_c = -2$~MHz from the respective cavity mode. Lines are guides to the eye.}
\label{ramprate}
\end{figure}

\begin{figure}[t]
 \includegraphics[width=0.75\textwidth]{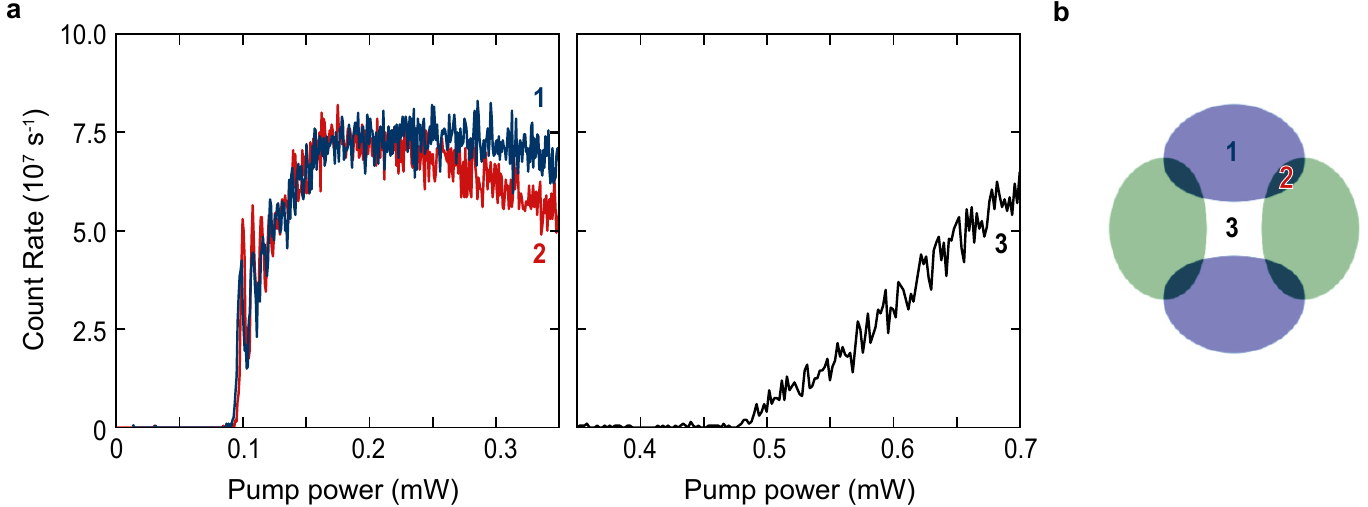}
\caption{\textbf{Threshold versus position in $l+m=1$ family.}  \textbf{a},  Cavity transmission, measured on a single photon counter, versus transverse pump power for three positions in the transverse plane of the cavity.  Measurement at detuning $\Delta_c =-28.8$~MHz  from the TEM$_{01}$ mode.  The three positions are shown in \textbf{b}: 1 (dark blue) at the antinode of TEM$_{10}$, 2 (red) in between modes, and 3 (black) at the node at center of the cavity. Onset of superradiance occurs at the same threshold for positions 1 and 2, while threshold for position 3 is much less sharp and at higher pump power, illustrating the poorer coupling when the BEC is at a node.  See Extended Data Fig.~\ref{ramprate} for more information about the dynamics of the organization transition at the node.}
\label{threshold}
\end{figure}

\begin{figure}[t]
 \includegraphics[width=0.60\textwidth]{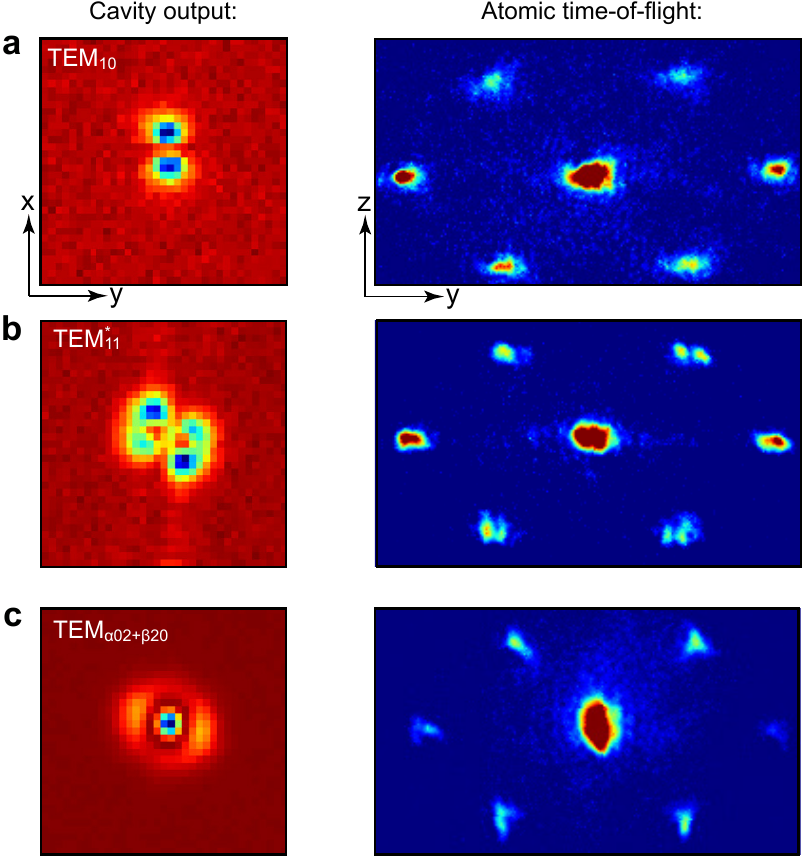}
\caption{\textbf{Structure factor measurement for supermode-DW-polariton condensates in $l+m=1$ and $2$ families.}  
 \textbf{a},  The nodal structure factor is not apparent for the mode orthogonal the TEM$_{01}$ mode shown in Fig.~\ref{structurefactor}d, because for imaging along $\hat{x}$, the node in atomic density along $\hat{y}$ is obscured by the column  integration inherent in the absorption imaging process. 
\textbf{b},  Image of  photonic component of a supermode-DW-polariton condensate that is mostly TEM$_{11}$, with only a few percent admixture of  TEM$_{02}$ and  TEM$_{20}$.  A nodal structure factor is evident in the first-order Bragg peaks  for this  even-parity supermode-DW-polariton condensate, because the BEC sits at a cavity nodal plane oriented parallel to the atomic absorption imaging axis $\hat{x}$.  \textbf{c}, By contrast, no nodal structure factor is evident for this nearly azimuthally symmetric supermode because 1) most atoms are located in the central antinode and 2) the  nodal plane parallel to $\hat{x}$ is obscured by the  atomic density in the ring  parallel to $\hat{y}$; $\alpha\approx81$\% and  $\beta\approx19$\%.  The high-order fringes along $\hat{x}$ in panel \textbf{b} are, we believe, the admixture  into the supermode-DW-polariton condensate of a very high-order transverse mode or set of modes from a family of modes  originating from half a free-spectral range lower in frequency.  While such modes are not common, we do find them at very specific $\Delta_c$'s.} 
\label{highorderSF}
\end{figure}

\begin{figure}[t]
 \includegraphics[width=0.60\textwidth]{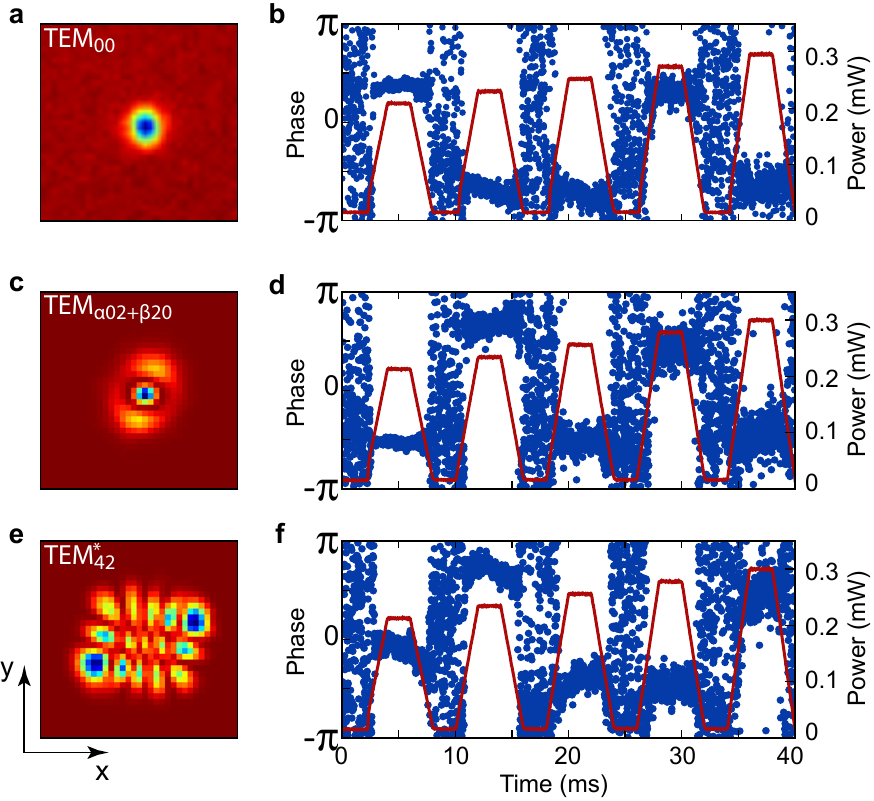}
\caption{\textbf{Heterodyne measurement of the phase locking of the supermode-DW-polariton condensate.}  \textbf{a},  \textbf{c}, and \textbf{e}, Images of the photonic component of the supermode-DW-polariton condensates for which phase measurements are made.   The supermode in \textbf{c}, also shown as  Fig.~\ref{decomposition}d.i, is a mixture of $\alpha\approx81$\% TEM$_{02}$ and  $\beta\approx19$\% TEM$_{20}$, while the supermode in \textbf{e} is dominated by the TEM$_{42}$  mode.  \textbf{b},  \textbf{d}, and \textbf{f}, Heterodyne measurements of the phase of the supermode-DW-polariton condensates in panels  \textbf{a},  \textbf{c}, and \textbf{e}.   The red overlay shows the power of the transverse pump versus time as the atoms self-organize and melt five times in the single experimental cycle.  The supermode-DW-polariton condensate picks one of two possible phases, separated by $\sim$$\pi$, with respect to the transverse pump for each organization event. This is a signature of the $Z_{2}$ symmetry being broken upon condensation:  the locking of phase corresponds to the two possible atomic checkerboard patterns~\cite{Domokos:2002bv,Black:2003bd,Asboth:2005ho}.  \textbf{c}--\textbf{f}, This phase locking is evident even for supermode-DW-polariton condensates with higher-order superpositions of the cavity supermodes.}
\label{phase}
\end{figure}

\end{document}